\documentclass[twocolumn,showpacs,preprintnumbers,nofootinbib,amsmath,amssymb]{revtex4}

\usepackage{graphicx}
\usepackage{amsfonts,latexsym}
\usepackage{euscript,amssymb}
\usepackage{epsfig}
\usepackage{dcolumn}
\usepackage{bm}

\catcode`\@=11
\def\Let@{\relax\iffalse{\fi\let\\=\cr\iffalse}\fi}
\def\vspace@{\def\vspace##1{\crcr\noalign{\vskip##1\relax}}}
\def\multilimits@{\bgroup\vspace@\Let@
 \baselineskip\fontdimen10 \scriptfont\tw@
 \advance\baselineskip\fontdimen12 \scriptfont\tw@
 \lineskip\thr@@\fontdimen8 \scriptfont\thr@@
 \lineskiplimit\lineskip
 \vbox\bgroup\ialign\bgroup\hfil$\m@th\scriptstyle{##}$\hfil\crcr}

\def\Sb{_\multilimits@}
\def\endSb{\crcr\egroup\egroup\egroup}
\def\Sp{^\multilimits@}

\newcommand{\nn}{\nonumber}
\newcommand{\be}[1]{\begin{equation}\label{#1}}
\newcommand{\ee}{\end{equation}}
\newcommand{\ba}[1]{\begin{eqnarray}\label{#1}}
\newcommand{\ea}{\end{eqnarray}}
\newcommand{\rf}[1]{(\ref{#1})}

\newcommand{\sign}{ \mbox{\rm sign}\,}

\begin{document}

\title{$1/R$ multidimensional gravity with form-fields: stabilization
of extra dimensions, cosmic acceleration and domain walls}


\author{Tamerlan Saidov}\email{tamerlan-saidov@yandex.ru}
\author{Alexander Zhuk} \email{zhuk@paco.net}
\affiliation{Department of Theoretical Physics and Astronomical Observatory,\\
Odessa National University,\\
2 Dvoryanskaya Street, 65026 Odessa, Ukraine
 }


\begin{abstract}
We study multidimensional gravitational models with scalar
curvature nonlinearity of the type $1/R$ and with form-fields
(fluxes) as a matter source. It is assumed that the higher
dimensional space-time undergoes Freund-Rubin-like spontaneous
compactification to a warped product manifold. It is shown that
for certain parameter regions the model allows for a freezing
stabilization of the internal space near the positive minimum of
the effective potential which plays the role of the positive
cosmological constant. This cosmological constant provides the
observable late-time accelerating expansion of the Universe if
parameters of the model is fine tuned. Additionally, the effective
potential has the saddle point. It results in domain walls in the
Universe. We show that these domain walls do not undergo
inflation.
\end{abstract}


\maketitle


\section{\label{intro}Introduction}
\setcounter{equation}{0}

There are two great challenges in modern theoretical physics and
cosmology.  The first big puzzle consists in a "dark side" of our
Universe. Recent observations of the luminosity distances of type
Ia supernovas (SNIa), CMB angular temperature fluctuations on
degree scales, and measurements of the power spectrum of galaxy
clustering indicate (see e.g. \cite{WMAP2006}) that our Universe
spatially flat with $\sim 23\%$ of its critical energy in
non-relativistic cold dark matter and $\sim 73\%$ in a smooth
component having a large negative pressure (dark energy). The
latter one results in accelerating expansion of our Universe which
began approximately at redshift $z\sim 1$ and continues until
present time. On the other hand, there is also possibility that
the late-time accelerating expansion of our Universe is caused by
modification of gravity on Galactic scales. For example, it was
proposed \cite{R^{-1}} to add a $1/R$ term in the Einstein-Hilbert
action to modify General Relativity\footnote{Terms with negative
powers of curvature can originate due to compactification of some
fundamental string/M-theory \cite{new}.}. It is clear that such
modification may affect dynamics of the Universe at late times of
its evolution and on large scales where the scalar curvature
becomes small. In fact, it was shown (see e.g. \cite{Carroll})
that this term can provide accelerating expansion of the Universe
without the need of introducing dark energy.

The second great challenge is possible multidimensionality of our
Universe which naturally follows from theories unifying different
fundamental interactions with gravity, such as string/M-theory.
So, there is big temptation to explain the dark matter and the
accelerating expansion of our Universe with the help of extra
dimensions. However, it is well known that dynamical behavior of
internal spaces usually results in variations of the effective
four-dimensional fundamental "constants" (e.g. gravitational
constant, fine structure constant, etc.) (see e.g.
\cite{Zhuk(IJMP)}-\cite{Alimi}) and references therein). There are
strong experimental bounds on such variations \cite{Uzan}. So, one
of the main problems of higher-dimensional models lies in stable
compactification of the internal spaces. Scale factors of the
internal spaces play the role of scalar fields moving in our
four-dimensional space-time. Their dynamics is defined by an
effective potential in dimensionally reduced theory. Thus, the
internal spaces are stabilized in the case of a minimum of this
potential \cite{GZ(PRD1997)}. Small excitations around this
minimum look in our Universe as massive scalar fields
(gravitational excitons/radions \cite{GZ(PRD1997)}) with Planck
scale suppression of their interaction with usual matter
\cite{GSZ}. Therefore they may play the role of dark matter.
Additionally, if the minimum of the effective potential is
positive, it contributes in positive cosmological constant
providing acceleration of the Universe.

In the present paper, we consider nonlinear gravitational
multidimensional cosmological model with action of the type
$R+1/R$ with form-fields as a matter source. We also include a
bare cosmological term as an additional parameter of the theory.
It is assumed that the corresponding higher-dimensional space-time
manifold undergos a spontaneous compactification to a manifold
with warped product structure of the external and internal spaces.
Each of spaces has its own scale factor. A model without
form-fields and bare cosmological constant was considered in paper
\cite{GZBR} where the internal space freezing stabilization was
achieved due to negative minimum of the effective potential. Thus,
such model is asymptotically AdS without accelerating behavior of
our Universe. It is well known that inclusion of usual matter can
uplift potential to the positive values \cite{GMZ(PRDb)}. One of
the main task of our present investigations is to observe such
uplifting due to the form-fields. Indeed, we demonstrate that for
certain parameter regions the late-time acceleration scenario in
our model becomes reachable. However, it is not simple uplifting
of the negative minimum of the theory \cite{GZBR} to the positive
values. The presence of the form-fields results in much more rich
structure of the effective potential then in \cite{GZBR}. Here, we
obtain additional branches with extremum points, and one of such
extremum corresponds to the positive minimum of the effective
potential. This minimum plays the role of the positive
cosmological constant. With the corresponding fine tuning of the
parameters, it can provide the late-time accelerating expansion of
the Universe. Moreover, we show that for this branch of the
effective potential there is also a saddle point. Thus, we obtain
domain walls which separate regions with different vacua. We
demonstrate that these domain walls do not undergo inflation
because our effective potential is not flat enough around the
saddle point.

It is also worth of noting that the effective potential in our
reduced model has a branchpoint. It gives very interesting
possibility to investigate transitions from one branch to another
by analogy with catastrophe theory or similar to phase transitions
in statistical theory. This idea needs more detail investigation
in a separate paper.

The paper is structured as follows. In section \ref{setup} we
present a brief description of multidimensional models with scalar
curvature nonlinearity $f(R)$ and the form-fields as a matter
source. Then, we perform dimensional reduction and obtain
effective four-dimensional action with effective potential.
General formulas from this section are applied to our specific
model $f({R})={R}-\mu /{R}-2\Lambda_{D}$ in section \ref{model}.
Here, we obtain the effective potential minimum conditions. These
conditions are analyzed in sections \ref{zero Lambda} and
\ref{positive Lambda} for the cases of zero and positive effective
cosmological constants respectively. Furthermore, in section
\ref{acceleration} we demonstrate that the positive minimum of the
effective potential plays the role of the positive cosmological
constant and can provide the late-time accelerating expansion.
Additionally, this minimum is accompanied by a saddle point. It
results in non-inflating domain walls in the Universe. The main
results are summarized and discussed in the concluding section
\ref{conclu}.


\section{\label{setup}General setup}
\setcounter{equation}{0}

We consider a $D= (D_0+D^{\prime})-$dimensional nonlinear
gravitational theory with action functional
\ba{1.1} S&=&\frac{1}{2
\kappa^{2}_{D}}\int_{M}d^{D}x\sqrt{|\bar{g}|}
f(\bar{R})\nn\\&-&\frac{1}{2}\int_{M}d^{D}x\sqrt{|\bar{g}|}\sum^{n}_{i=1}
\frac{1}{d_{i}!}\left(F^{(i)}\right) ^{2}\, ,\ea
where $f(\bar R)$ is an arbitrary smooth function of a scalar
curvature $\bar R := R[\bar g]$ constructed from the
$D-$dimensional metric $\bar g_{ab}\; (a,b = 1,\ldots,D)$.
$D^{\prime}$ is the number of extra dimensions. $\kappa^2_D $
denotes the $D-$dimensional gravitational constant. In action
(\ref{1.1}), a form field (flux) $F$ has block-orthogonal
structure consisting of $n$ blocks. Each of these blocks is
described by its own antisymmetric tensor field $F^{(i)}
(i=1,\ldots ,n)$ of rank $d_{i}$ ($d_{i}$-form field strength).
Additionally, we assume that for the sum of the ranks holds
$\sum^{n}_{i=1}d_{i}=D'$.

Following Refs. \cite{GZBR}-\cite{GMZ(ASS)}, we can show that the
nonlinear gravitational theory \rf{1.1} is equivalent to a linear
theory $R = R[g]$ with conformally transformed metric
\be{1.2} g_{ab} = \Omega^2 \bar g_{ab} = \left[ f'(\bar
R)\right]^{2/(D-2)}\bar g_{ab}\;  \ee
and an additional minimal scalar field $\phi=\ln[f'(\bar R)]/A$
coupled with fluxes. The scalar field $\phi$ is the result and the
carrier of the curvature nonlinearity of the original theory.
Thus, for brevity, we shall refer to the field $\phi$ as
nonlinearity scalar field. A self-interaction potential $U(\phi )$
of the scalar field $\phi$ reads
\be{1.3} U(\phi ) = \frac12 e^{- B \phi} \left[\; \bar R (\phi
)e^{A \phi } - f\left( \bar R (\phi )\right) \right]\; , \ee
where
\be{1.4} A = \sqrt{\frac{D-2}{D-1}}\, , \quad B = \frac
{D}{\sqrt{(D-2)(D-1)}}\, . \ee

Furthermore, we assume that the multidimensional space-time
manifold undergoes a spontaneous compactification \be{1.5} M
\longrightarrow M = M_0 \times M_1 \times \ldots \times M_n \ee in
accordance with the block-orthogonal structure of the field
strength $F$, and that the form fields $F^{(i)}$, each nested in
its own $d_i-$dimensional factor space $M_i\, (i=1,\ldots ,n)$,
respect a generalized Freund-Rubin ansatz \cite{FR}. Here,
($D_0=4$)-dimensional space-time $M_0$ is treated as our external
Universe with metric $g^{(0)}(x)$.

This allows us to perform a dimensional reduction of our model
along the lines of Refs.
\cite{GZ(PRD1997)}-\cite{GMZ(PRDa)},\cite{GZ(PRD2000)},\cite{RZ}.
The factor spaces $M_i$ are then Einstein spaces with metrics
$g^{(i)} \equiv e^{2\beta^i(x)} \gamma^{(i)}$ which depend only
through the warp factors $a_i(x) :=e^{\beta^i(x)}$ on the
coordinates $x$ of the external space-time $M_0$. For the
corresponding scalar curvatures holds $R\left[ \gamma^{(i)}\right]
=\lambda ^id_i\equiv r_i $ (in the case of the constant curvature
spaces $\lambda^i =k_i(d_i-1),\, \, k_i = 0,\pm 1$). The warped
product of Einstein spaces leads to a scalar curvature $\bar{R}$
which depends only on the coordinate $x$ of the $D_0-$dimensional
external space-time $M_0$: $\bar{R}[\bar{g}] = \bar{R}(x)$. This
implies that the nonlinearity field $\phi$ is also a function only
of $x$: $\phi = \phi (x)$. Additionally, it can be easily seen
\cite{GZBR} that the generalized Freund-Rubin ansatz results in
the following expression for the form-fields: $
\left(F^{(i)}\right)^2=f_i^2/a_i^{2d_i} $ where $ f^i=const $.

In general, the model will allow for several stable scale factor
configurations (minima in the landscape over the space of volume
moduli). We choose one of them (which we expect to correspond to
current evolution stage of our observable Universe), denote the
corresponding scale factors as $\beta^i_0$, and work further on
with the deviations $\hat \beta^i (x)= \beta^{i}(x) -
\beta^{i}_0$.

Without loss of generality\footnote{The difference between a
general model with $n>1$ internal spaces and the particular one
with $n=1$ consists in an additional diagonalization of the
geometrical moduli excitations. \label{n=1}}, we shall consider a
model with only one $d_1$-dimensional internal space. After
dimensional reduction and subsequent conformal transformation to
the Einstein frame the action functional \rf{1.1}
reads\footnote{The equivalency between original higher dimensional
and effective dimensionally reduced models was investigated in a
number of papers (see e.g.\cite{Ivashchuk}). The origin of this
equivalence results from high symmetry of considered models (i.e.
because of specific metric ansatz which is defined on the manifold
consisting of direct product of the Einstein spaces).}
\ba{1.6} S&=&\frac 1{2\kappa _0^2}\int\limits_{M_0}d^{D_0}x\sqrt{|
\tilde{g}^{(0)}|}\left\{ R\left[  \tilde{g}^{(0)}\right] -
\tilde{g}^{(0) \mu \nu}
\partial_{\mu}\varphi \partial_{\nu} \varphi \right.\nn\\ &-& \left. \tilde{g}^{(0) \mu \nu}
\partial_{\mu}\phi \partial_{\nu} \phi -2U_{eff} (\varphi ,\phi )
\right\} \, , \ea
where $\varphi := -\sqrt{d_1(D-2)/(D_0-2)}\, \hat \beta^1$ and
$\kappa^2_0 := \kappa^2_D/V_{d_1}$ denotes the $D_0-$dimensional
(four-dimensional) gravitational constant. $V_{d_1} \sim
\exp{(d_1\beta_{0}^{1})}$ is the volume of the internal space at
the present time.

A stable compactification of the internal space $M_1$ is  ensured
when its scale factor $\varphi$ is frozen at the minimum of the
effective potential
\be{1.7} U_{eff}=e^{b\varphi}\left[-\frac 12 R_1
e^{a\varphi}+U(\phi)+h e^{c\phi}e^{ad_1 \varphi} \right]\, , \ee
where $R_1 := r_1 \exp{(-2\beta^1_0)}$ defines the curvature of
the internal space at the present time and contribution of the
form-field into the effective action is described by
$h:=\kappa^2_D\,f^2_1 \exp{(-2d_1\beta^1_0)}
>0$. For brevity we introduce notations
\ba{1.8} a:&=&2\sqrt{\frac{D_0-2}{d_1(D-2)}}\; ,\;
b:\;=2\sqrt{\frac{d_1}{(D-2)(D_0-2)}}\; ,\nn\\ c:&=&
\frac{2d_1-D}{\sqrt{(D-1)(D-2)}}\; . \ea


\section{\label{model}The model}

\setcounter{equation}{0}

In this section we analyze the conditions of the compactification
for a model with
\be{2.1} f(\bar{R})=\bar{R}-\frac{\mu}{\bar{R}}-2\Lambda_{D}\;.
\ee
Then from the relation $f^{\prime}(\bar R) = \exp{(A\phi)}$ we
obtain
\be{R} \bar{R}=q\sqrt{\frac{|\mu |}{s(e^{A\phi}-1)}}\;,\quad
q=\pm1\;, \, s =\sign (\mu)\; . \ee
Thus, the ranges of variation of $\phi$ are $\phi\in(-\infty,0)$
for $\mu <0\; (s=-1)$ and $\phi\in(0,+\infty)$ for $\mu >0\;
(s=+1)$.

It is worth of noting that the limit $\phi \to \pm 0\; (f^{\prime}
\to 1 )$ corresponds to the transition to a linear theory: $f(\bar
R) \to \bar R -2\Lambda_D$ and $R\to \bar R$. This is general
feature of all nonlinear models $f(\bar R)$. For example, in our
case \rf{2.1} we obtain  $f(\bar R) = \bar R \left(2 - \exp (A\phi
)\right) - 2\Lambda_D \ \longrightarrow \bar R -2\Lambda_D$ for
$\phi \to 0$. {}From other hand, for particular model \rf{2.1},
eq. \rf{R} shows that the point $\phi = 0$ maps into infinity
$\bar R,R = \pm \infty$. Thus, in this sense, we shall refer to
the point $\phi =0$ as singularity.

For our model \rf{2.1}, potential \rf{1.3} $U(\phi)$ reads
\begin{equation}\label{U}
U(\phi)=\frac{1}{2}e^{-B\phi}\left(2qs\sqrt{|\mu|}\sqrt{se^{A\phi}-s}+
2\Lambda_{D}\right)\, .
\end{equation}

It is well known (see e.g.
\cite{GMZ(PRDb)},\cite{GMZ(PRDa)},\cite{GZ(PRD2000)}) that in
order to ensure a stabilization and asymptotical freezing of the
internal space $M_{1}$, the effective potential \rf{1.7} should
have a minimum with respect to both scalar fields $\varphi$ and
$\phi$. We remind that we choose the minimum position with respect
to $\varphi$ at $\varphi =0$. Additionally, the eigenvalues of the
mass matrix of the coupled $(\varphi,\phi)$-field system, i.e. the
Hessian of the effective potential at the minimum position,
\begin{equation}\label{Hessian}
    J:=\left.\left(
\begin{array}{cc}
  \partial^{2}_{\varphi\varphi}U_{eff} &  \partial^{2}_{\varphi\phi}U_{eff} \\\\
   \partial^{2}_{\phi\varphi}U_{eff} &  \partial^{2}_{\phi\phi}U_{eff} \\
\end{array}
\right)\right|_{extr}
\end{equation}
should be positive definite (this condition ensures the
positiveness of the mass squared of scalar field excitations).
According to the Silvester criterion this is equivalent to the
condition:
\begin{equation}\label{con}
J_{11}>0\;,\quad J_{22}>0\;,\quad \mbox{det}(J)>0\quad.
\end{equation}

It is convenient in further consideration to introduce the
following notations:
\begin{equation}\label{xc}
\phi_{0}:=\phi|_{extr}\,,\, X:=\sqrt{se^{A\phi_{0}}-s}>0\,
\rightarrow \, X_{(s=-1)}<1\, .
\end{equation}
Then we can rewrite potentials $U(\phi )$, $U_{eff}(\varphi ,\phi
)$ and derivatives of the $U_{eff}$ at an extremum (possible
minimum) position ($\varphi=0,\phi_{0}$) as follows:
\be{U0} U_{0}\equiv
U|_{extr}=\left(1+sX^{2}\right)^{-B/A}\left(qs\sqrt{|\mu|}X+
\Lambda_{D}\right), \ee
\be{Ueff0}
U_{eff}|_{extr}=-\frac{1}{2}R_{1}+U_{0}(X)+h\left(1+sX^{2}\right)^{c/A}\;,
\ee
\ba{vpUeff}
\partial_{\varphi}U_{eff}|_{extr}&=&-\frac{a+b}{2}R_{1}+bU_{0}(X)\nn\\&+&(ad_{1}+
b)h\left(1+sX^{2}\right)^{c/A}=0,\;\quad \ea
\ba{pUeff}
\partial_{\phi}U_{eff}|_{extr}&=&
ch\left(1+sX^{2}\right)^{c/A}-BU_{0}(X)\nn\\&+&\frac{qA\sqrt{|\mu|}}{2X}
\left(1+sX^{2}\right)^{1-B/A}=0,\qquad \ea
\ba{vpvpUeff}
\partial^{2}_{\varphi\varphi}U_{eff}|_{extr}&=&-\frac{(a+b)^2}{2}R_{1}+
b^{2}U_{0}(X)\nn\\&+&(ad_{1}+b)^{2}h\left(1+sX^{2}\right)^{c/A},\qquad
\ea
\be{pvpUeff}
\partial^{2}_{\varphi\phi}U_{eff}|_{extr}=chad_{1}\left(1+sX^{2}\right)^{c/A}\;,
\ee
\ba{ppUeff}
\partial^{2}_{\phi\phi}U_{eff}|_{extr}&=&ch\left(c-A+2B\right)
\left(1+sX^{2}\right)^{c/A}\nn\\&+&B(A-B)U_{0}(X)\nn \\
&-&
\frac{qs\sqrt{|\mu|}A^{2}}{4X^{3}}\left(1+sX^{2}\right)^{2-B/A}.
\:\qquad \ea
The most natural strategy for extracting detailed information
about the location of stability region of parameters in which
compactification is possible would consist in solving
(\ref{pUeff}) for $X$ with subsequent back-substitution of the
found roots into the inequalities (\ref{con}) and the equation
(\ref{vpUeff}). To get the main features  of the model under
consideration, it is sufficient to investigate two particular
nontrivial situations. Both of these cases are easy to handle
analytically.

\section{\label{zero Lambda}Zero effective cosmological constant: $\Lambda_{eff}=0$}

\setcounter{equation}{0}

It can be easily seen from eqs. \rf{Ueff0} and \rf{vpUeff} that
condition $\Lambda_{eff}=U_{eff}|_{extr}=0$ results in relations
\be{R1}
R_{1}=2d_{1}h\left(1+sX^{2}\right)^{c/A}=\frac{2d_{1}}{d_{1}-1}U_{0}(X)\;,\quad
d_{1}\geq2\; , \ee
which enable us to get from eq. (\ref{pUeff}) quadratic equation
for $X$
\be{4.1} (d_1+1)X^2 + qsd_1zX - s(d_1-1) = 0\, ,\quad z\equiv
2\Lambda_{D}/\sqrt{|\mu|} \ee
with the following solutions:
\ba{s1}
X_{p}&=&qs\frac{d_{1}}{2(d_{1}+1)}\left(-z+p\:\sqrt{z^2+4s\frac{d^{2}_{1}
-1}{d_{1}^{2}}}\right)\; ,\nn\\ p&=&\pm 1\,.\ea
In the case $s=-1\;$ parameter $z$ should satisfy condition $
|z|\geq z_{0}\equiv2\sqrt{d_{1}^{2}-1}/d_{1}<2$ and for $z=z_0$
two solutions $X_{p}$ degenerate into one: $X_{p} \equiv X_0 =
-qs\sqrt{(d_1-1)/(d_1+1)}$.

Because of conditions $h\geq0$ and $e^{A\phi_{0}} = 1+sX^2
>0$, the relations \rf{R1} show that  parameters $R_{1}$ and
$U_{0}(X)$ should be non-negative: $R_{1}\geq0 , \;
U_{0}(X)\geq0$. Obviously, only one of the solutions \rf{s1}
corresponds to a minimum of the effective potential. With respect
to this solution we define parameters in the relation \rf{R1}.
Therefore, we must distinguish now which of $X_{p}$ corresponds to
the minimum of $U_{eff}$. Let us investigate solutions \rf{s1} for
the purpose of their satisfactions to conditions $e^{A\phi_{0}}> 0
,\, U_{0}(X)\geq 0$ and $X_{p} \geq 0$.

\textbf{The condition $e^{A\phi_{0}}=1+sX_{p}^{2}>0$:}

Simple analysis shows that solutions $X_{p}$ satisfy  this
inequality for the following combinations of parameters:
\begin{equation}\label{Xs}
\begin{tabular}{|c|c|}
\hline $\begin{array}{c}\mu >0\\(s=+1)\end{array}$ & $\; \, p=\pm1
\; :
\; z\in(-\infty,+\infty)\quad\qquad\qquad$\\
\hline $\begin{array}{c}\mu<0\\(s=-1)\end{array}$ &
$\begin{array}{c}p=+1
\;:\; z\in(-2,-z_{0})\cup(z_{0},+\infty) \\
 p=-1 \;:\; z\in(-\infty,-z_{0})\cup(z_{0},2)\quad\:\\ z=z_{0}\qquad\qquad\qquad\qquad\qquad\quad\qquad\qquad\end{array}$ \\
\hline
\end{tabular}
\end{equation}

\textbf{The condition $U_{0}(X)\geq0$:}

As appears from eq. \rf{U0}, this condition takes place if $X_{p}$
satisfies inequality $2qsX_{p}+z\geq 0$ which leads to the
conditions:
\begin{equation}\label{val}
\begin{tabular}{|c|c|}
  \hline
      $\begin{array}{c}\mu >0\\(s=+1)\end{array}$ &
      $\begin{array}{c}\; \; p=+1\;:\;z\in(-\infty,+\infty)\;\end{array}$ \\
  \hline
   $\begin{array}{c}\mu<0\\(s=-1)\end{array}$ &
   $\begin{array}{c}p=+1\;  :\; z\in(z_{0},+\infty)\;\;\;\\
   p=-1\;  :\; z\in(z_{0},2]\qquad\:\\z=z_{0}\qquad\qquad\qquad\qquad\qquad\end{array}$ \\
  \hline
\end{tabular}
\end{equation}
\textbf{The condition $X_{p}>0$:}

This condition is satisfied for the combinations:
\begin{equation}\label{X<>}
\begin{tabular}{|c|c|}
  \hline
      $\begin{array}{c}\mu >0\\(s=+1)\end{array}$ &
      $\begin{array}{c}q=+1:\left\{\begin{array}{c}
      z<0:\quad X_{+}>0 \quad X_{-}<0\\
      z>0:\quad X_{+}>0 \quad X_{-}<0\end{array}\right.\\
      q=-1:\left\{\begin{array}{c}
      z<0:\quad X_{+}<0 \quad X_{-}>0\\
      z>0:\quad X_{+}<0 \quad X_{-}>0\end{array}\right.\end{array}$ \\
  \hline
   $\begin{array}{c}\mu <0\\(s=-1)\end{array}$ &
   $\begin{array}{c}q=+1:\left\{\begin{array}{c}
      z<0:\quad X_{+}<0 \quad X_{-}<0\\
      z>0:\quad X_{+}>0 \quad X_{-}>0\end{array}\right.\\
      q=-1:\left\{\begin{array}{c}
      z<0:\quad X_{+}>0 \quad X_{-}>0\\
      z>0:\quad X_{+}<0 \quad X_{-}<0\end{array}\right.\end{array}$ \\
  \hline
\end{tabular}
\end{equation}

The comparison of \rf{Xs}, \rf{val} and \rf{X<>} shows that they
are simultaneously satisfied only for the following combinations:
 \begin{equation}\label{XF}
 \begin{tabular}{|c|c|}
  \hline
      $\begin{array}{c}\mu >0\\(s=+1)\end{array}$ &
      $\begin{array}{c}\; \; \; \; p=+1\, :\;\;q=+1\;:\;z\in(-\infty,+\infty)\end{array}\;$ \\
  \hline
   $\begin{array}{c}\mu <0\\(s=-1)\end{array}$ &
   $\begin{array}{c}p=+1\;:\;\;q=+1\;:\;z\in(z_{0},+\infty)\\
\;\;p=-1\;:\;\;q=+1\;:\;z\in(z_{0},2)\qquad\:\\\;q=+1\;:\;z=z_{0}\qquad\qquad\qquad\qquad\qquad\end{array}$ \\
  \hline
\end{tabular}
\end{equation}

Additionally, the extremum solutions $X_{p}$ should correspond to
the minimum of $U_{eff}$. The inequalities (\ref{con}) are the
sufficient and necessary  conditions for that. We analyze them in
the case of four-dimensional external space $D_0=4$. Taking into
account definitions \rf{1.4}, \rf{1.8}, \rf{vpvpUeff}-\rf{ppUeff}
and relations \rf{R1}, for $J_{11}, J_{22}$ and $J_{21}$ we get
respectively:
\be{Lj11} J_{11}=\frac{8}{d_{1}+2}U_{0}(X_{p})\, , \ee
 \begin{equation}\label{Lj22}
 \begin{split}
  J_{22}=&-\sqrt{|\mu|}\left[\frac{6d_{1}(2qsX_{p}+z)\left(1+sX_{p}^{2}\right)^{-\frac{d_{1}+4}{d_{1}+2}}}{(d_{1}-1)(d_{1}+2)(d_{1}+3)}
  \right.
\\ &+\left.\frac{qs(d_{1}+2)}{4X^{3}_{p}(d_{1}+3)}
\left(1+sX_{p}^{2}\right)^{\frac{d_{1}}{d_{1}+2}}\right] \, ,
\end{split}
\end{equation}
\ba{Lj21}
J_{21}=&-&\sqrt{|\mu|}\left[\frac{(d_{1}-4)(2qsX_{p}+z)}{(d_{1}-1)(d_{1}+2)}\right.\nn\\
&\times &\left. \sqrt{\frac{2}{d_{1}(d_{1}+3)}}\left(1+sX_{p}^{2}
\right)^{-\frac{d_{1}+4}{d_{1}+2}}\right]\, . \quad\ea
We supposed in these equations that each of $X_{p}$ can define
zero minimum of $U_{eff}$. In what follows, we shall check this
assumption for every $X_{p}$ with corresponding combinations of
signs of the parameters $s$ and $q$ in accordance with the table
\rf{XF}.

According to the Silvester criterion \rf{con}, $J_{11}$ should be
positive. Thus eqs. \rf{R1} and \rf{Lj11} result in the following
conclusions: the potential $U_0$ should be positive $U_0>0$, the
internal space should have positive curvature $R_1>0$ (hence,
$d_1>1$) and its stabilization (with zero minimum
$\Lambda_{eff}=0$) takes place only in the present of form-field
($h>0$). Transition from the non-negativity condition $U_0\geq 0$
to the positivity one $U_0>0$ corresponds to the only substitution
in \rf{val} $(z_{0},2] \rightarrow (z_{0},2)$  for the case
$s=-1,\; p=-1$. Exactly this interval $(z_{0},2)$ appears in
concluding table \rf{XF}. Therefore, $J_{11}$ is positive for all
$X_{p}$ from the table \rf{XF}.

\begin{figure}[htbp]
\centerline{\includegraphics[width=3in,height=2in]{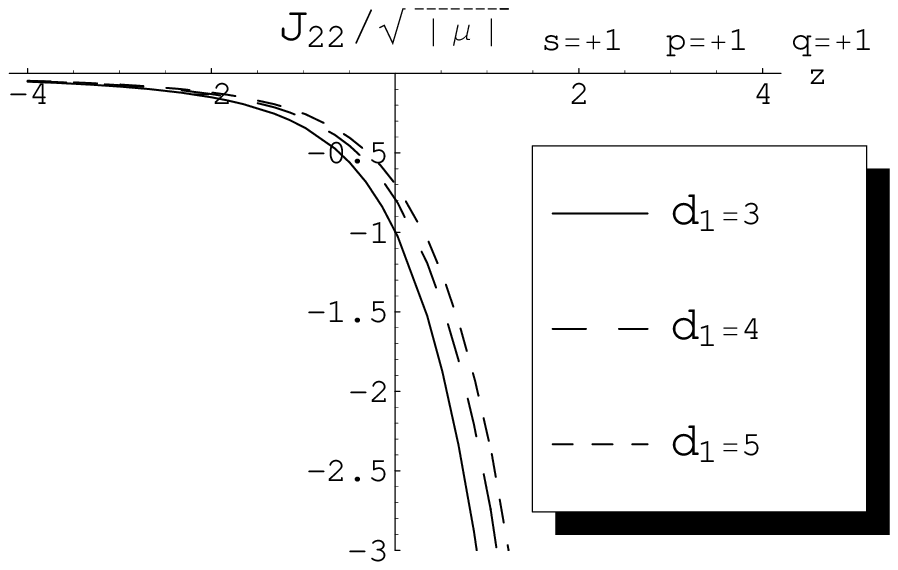}}
\centerline{\includegraphics[width=3in,height=2in]{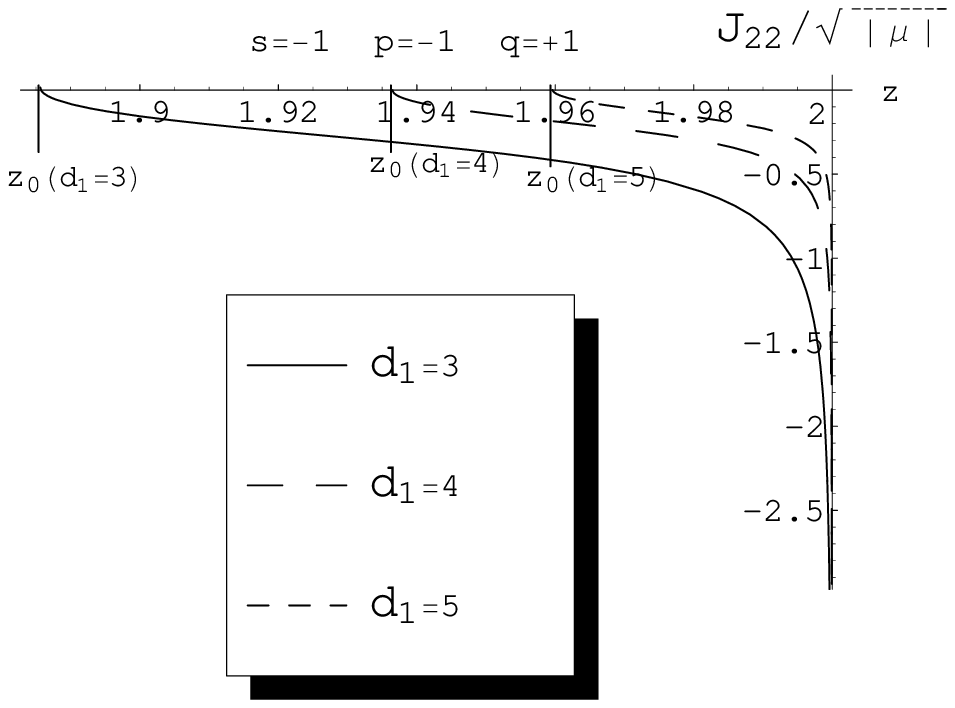}}
 \caption{Typical form of $J_{22}/\sqrt{|\mu|}$ (eq. \rf{Lj22}) for parameters
 $s=+1,\; q=+1,\; p=+1$ (upper) and $s=-1,\; q=+1,\; p=-1$ (lower). For both of these
 combinations of the parameters, $J_{22}<0$. \label{Lj22-1}}
\end{figure}

\begin{figure}[htbp]
\centerline{\includegraphics[width=3in,height=2in]{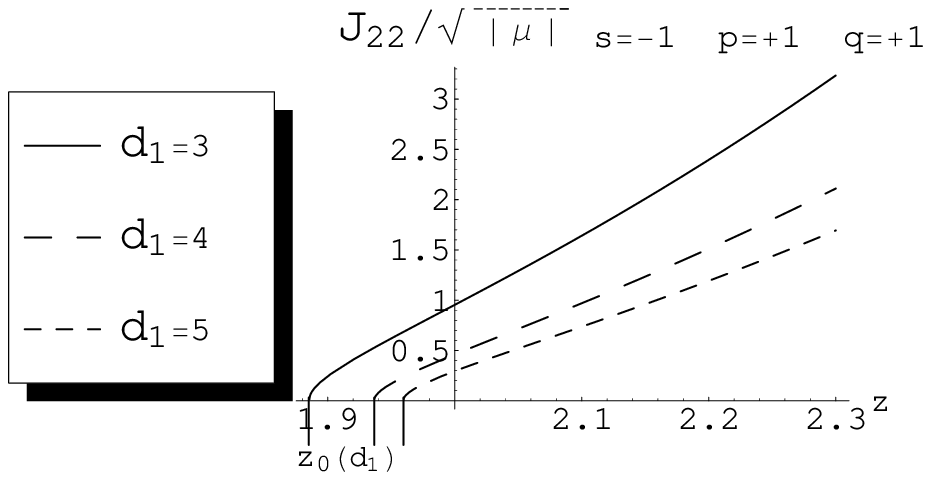}}
\centerline{\includegraphics[width=3in,height=2in]{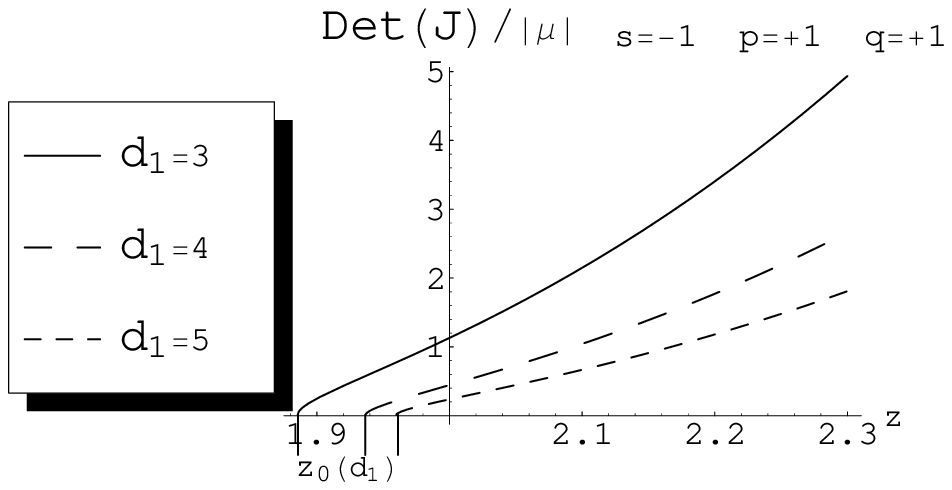}}
 \caption{Typical form of $J_{22}/\sqrt{|\mu|}$ (upper) and
 $\mbox{det}(J)/|\mu|$ (lower) (see eqs. \rf{Lj11}-\rf{Lj21}) for parameters
  $s=-1,\; q=+1,\; p=+1$. For this combination of the parameters, $J_{22}>0$
  and $\mbox{det}J>0$. \label{Lj22-2}}
\end{figure}

Concerning expressions $J_{22}$ and
$\mbox{det}(J)=J_{11}J_{22}-J_{12}^2$,\, graphical plotting (see
Fig.\ref{Lj22-1} and Fig.\ref{Lj22-2}) demonstrates that they are
negative for $s=+1,\; p=+1,\; q=+1$ and $s=-1,\; p=-1,\; q=+1$ but
positive in the case $s=-1,\; p=+1,\; q=+1$. For this latter
combination $z \in (z_0,\infty)$. The case $s=-1,\; q=+1$ and $z =
z_0$ should be investigated separately. Here, $X_{p}\equiv
X_{0}=\sqrt{(d_{1}-1)/(d_{1}+1)}$ and for $J_{22}$ and $J_{21}$ we
obtain:
\ba{J022}
J_{22}=&-&\sqrt{|\mu|}X^{-3}_{0}\left(1+sX_{0}^{2}\right)^{-\frac{d_{1}+4}{d_{1}+2}}
\nn\\
&\times
&\left(\frac{12(d_{1}-1)-(d_{1}+2)^{2}}{(d_{1}+1)^{2}(d_{1}+2)(d_{1}+3)}\right),
\quad\ea
\ba{J021}
J_{21}=&-&\sqrt{|\mu|}X^{-3}_{0}\left(1+sX_{0}^{2}\right)^{-\frac{d_{1}+4}{d_{1}+2}}
\nn\\ &\times
&\left(\frac{2\sqrt{2}(d_{1}-4)(d_{1}-1)}{(d_{1}+1)^{2}(d_{1}+2)\sqrt{d_1(d_{1}+3)}}\right).
\qquad \ea
It can be easily seen from eqs. (\ref{J022}) and \rf{J021} that
$J_{22}>0$ for $d_{1}\neq 4$ and   $J_{22} = J_{21}=0$ for
$d_{1}=4$. Additionally, $\mbox{det}(J)>0$ for $d_{1}\neq 4$.

Thus, we can finally conclude that zero minimum of the effective
potential $U_{eff}$ takes place either for $s=-1,\; q=+1,\; z\in
(z_{0},\infty), \forall \, d_{1}>1$ (position of this minimum is
defined by solution \rf{s1} with $p=+1$) or for $ s=-1,\; q=+1,\;
z=z_{0}, \forall \, d_{1}\neq 4$. Concerning the signs of
parameters, we obtain that $\mu <0$ and $\Lambda_D>0$.


\section{\label{positive Lambda}Decoupling of excitations: $d_{1}=D_{0}$}

\setcounter{equation}{0}

It can be easily seen from eq. \rf{1.8} that in the case
$d_{1}=D_{0}$ parameter $c=0$ that leads to condition $
\partial^{2}_{\varphi\phi}U_{eff}|_{extr} =0$ (see eq.
\rf{pvpUeff}. Thus the Hessian \rf{Hessian} is diagonalized. It
means that the excitations of the fields $\varphi$ and $\phi$ near
the extremum position are decoupled from each other\footnote{In
the vicinity of a minimum of the effective potential, squared
masses of these excitations are $m^2_{\varphi} =J_{11}$ and
$m^2_{\phi}=J_{22}$.}.

Dropping the $h$ term in eq. \rf{pUeff} (because of $c=0$) and
taking into account eq. \rf{U0}, we obtain quadratic equation for
$X$
\be{5.1} (D+2)X^2 + DqszX - (D-2)s =0\, , \ee
which for $D_0=d_1$ exactly coincides with eq. \rf{4.1}. Thus, in
spite of the fact that we does not use the condition
$\Lambda_{eff} =0$, we obtain in the case $d_1=D_0$ precisely the
same solutions \rf{s1}. However, parameters $R_1, U_0$ and $h$
satisfy now relations different from \rf{R1}. For example, for the
most physically interesting case $D_0=d_1=4$, eqs. \rf{vpUeff} and
\rf{Ueff0} result in the following relations:
\be{RR} R_{1}=4\left[\frac{1}{3}U_{0}(X)+h\right] \; , \quad
\Lambda_{eff}(X)=\frac{1}{3}U_{0}(X)-h\, . \ee Nonzero components
of the Hessian read
\ba{Dj11} J_{11}&=&\frac{2}{3}\left[9h-U_{0}(X)\right]\:,\:
\\J_{22}&=&
    -\sqrt{|\mu|}\left[\frac{4(z+2qsX_{p})}{21(1+sX_{p}^{2})^{4/3}}+\frac{3qs(1+sX_{p}^{2})^{2/3}}{14X_{p}^{3}}\right].
    \nn
\ea
In this section we are looking for a positive minimum of the
effective potential. It means that $\Lambda_{eff} >0$, $J_{11}>0$
and $J_{22}>0$. From the positivity of $J_{11}$ and
$\Lambda_{eff}$ we obtain respectively\footnote{It is interesting
to note that (in the case $D_0=d_1=4$) relations \rf{5.1},
$J_{11}$ in \rf{Dj11} and inequalities \rf{J11>0}, \rf{posit}
coincide with the analogous expressions in paper \cite{GMZ(PRDb)}
with quadratic nonlinear model. This is not surprising because
they do not depend on the form of nonlinearity $f(\bar R)$ (and,
consequently, on the form of $U(\phi )$). However, the expressions
for $J_{22}$ are different because here we use the exact form of
$U(\phi )$.}:
\be{J11>0} J_{11}>0 : \quad
16h>R_{1}>16U_{0}(X)/9>8\Lambda_{eff}\; \ee
and
\be{posit} \Lambda_{eff}>0 : \quad h>R_{1}/16>U_{0}(X)/9>h/3>0\;.
\ee
These inequalities show that for the considered model positive
minimum of the effective potential is possible only in the case of
positive curvature of the internal space $R_1>0$ and in the
presence of the form field $(h>0)$.

To realize which combination of parameters $s, p$ and $q$ ensures
the minimum of the effective potential, we should perform analysis
as in the previous case with $\Lambda_{eff}=0$. However, there is
no need to perform such analysis here because solutions of eq.
\rf{5.1} coincides with \rf{s1} and all conditions for $X_{p}$ and
$U_0$ are the same as in the previous section. Thus, we obtain
concluding table of the form \rf{XF}. Additionally, it can be
easily seen that expressions of $J_{22}$ in \rf{Dj11} and
\rf{Lj22} exactly coincide with each other if we put $d_1=4$ in
the latter equation\footnote{It follows from the fact that in
\rf{Lj22} we already put $D_0=4$. Although we use in this equation
the relation \rf{R1} between $h$ and $U_0$, it enters here in the
combination which is proportional to $c$. Thus, this combination
does not contribute if we put additionally $d_1=4$.}. Hence, we
can use Fig.\ref{Lj22-1} and Fig.\ref{Lj22-2} (for the lines with
$d_1=4$) to analyze the sign of $J_{22}$. With the help of these
pictures as well as keeping in the mind that $J_{22}(d_1=4, z=z_0)
= 0$, we obtain that the only combination which ensures the
positive minimum of $U_{eff}$ is: $s=-1\, ,\; p=+1\, ,\; q=+1\,$
and $z\in(z_{0}=1.936,+\infty)$. It is clear that potential
$U_0(X)$ in eqs. \rf{RR}-\rf{posit} is defined by solution of eq.
\rf{5.1} (i.e. \rf{s1} for $d_1=4$) with this combination of the
parameters. Because $s=-1$ and $z>0$ , the parameters $\mu$ and
$\Lambda_D$ should have the following signs: $\mu<0$ and
$\Lambda_D>0$.

Additionally, it is easy to verify that second solution of eq.
\rf{5.1} $X_-$ (with $p=-1$ and $s=-1, \, q=+1$) does not
correspond to the maximum of the effective potential $U_{eff}$.
Indeed, we have here $\partial_{\phi} U_{eff}(X_{-})=0$ but
$\partial_{\varphi} U_{eff}(X_{-})\ne 0$.

\begin{figure}[htbp]
\centerline{\includegraphics[width=3.5in,height=3in]{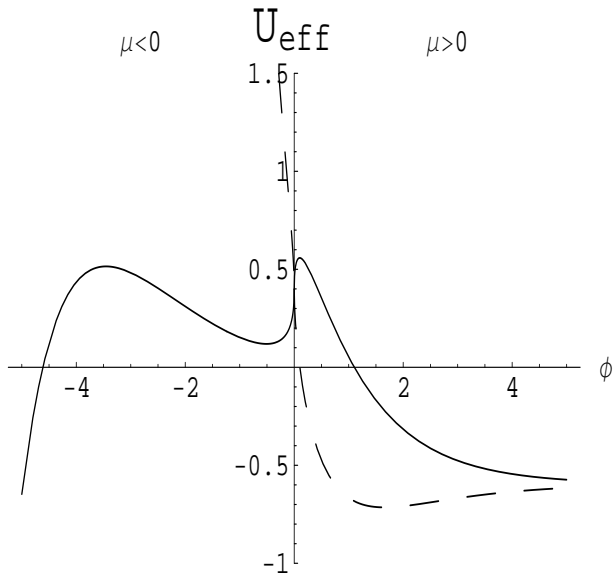}}
\caption{ Profile $\varphi=0$ of the effective potential
$U_{eff}(\varphi ,\phi)$ for parameters
$z=1.99\;(\Lambda_{D}=1.99/2\;,\;|\mu|=1),\;D_{0}=d_{1}=4$ and
$h=U_{0}/6$. The rest of parameters can be found from relations
\rf{RR}. Solid and dashed lines describe branches $q=+1$ and
$q=-1$ respectively. For the solid line, there is only one local
minimum of $U_{eff}$  which is defined by solution $X_{p}$ with
the following parameters: $s=-1,\; p=+1,\; q=+1$. Left ($s=-1,\;
p=-1,\; q=+1$) and right ($s=+1,\; p=+1,\; q=+1$) maxima are not
extrema of $U_{eff}$ because in these points $\partial_{\phi}
U_{eff}=0$ but $\partial_{\varphi} U_{eff}\ne 0$. Analogously, for
right dashed line the solution $X_{p}$ (with $s=+1,\; p=-1,\;
q=-1$) does not correspond to the extremum of $U_{eff}$.
\label{pot}}
\end{figure}

Fig. \ref{pot} demonstrates the typical profile $\varphi =0$ of
the effective potential $U_{eff}(\varphi ,\phi)$ in the case of
positive minimum of $U_{eff}$ considered in the present section.
This picture is in good concordance with the table \rf{XF}.
According to this table, positive extrema of $U_{eff}$ are
possible only for the branch $q=+1$ of the solution \rf{R} (solid
lines in Fig. \ref{pot}). We see that for $z_0<z<2$ we can have 3
such extrema: one for positive $\mu>0 \to s=+1$ and two for
negative $\mu<0 \to s=-1$. Our investigations show that in the
left half plane (i.e. for $\mu<0$) the right extremum ($p=+1$ in
eq. \rf{s1}) is the local minimum\footnote{It can be easily seen
for the branch $q=+1,\; s=-1$ that $U(\phi \to -\infty ) \to
+\infty$ for $z\geq 2$ and $U(\phi \to -\infty ) \to -\infty$ for
$z< 2$. Thus, for $z\geq 2$ this minimum becomes global one.} and
the left maximum ($p=-1$) is not the extremum of $U_{eff}$ because
here $\partial_{\varphi} U_{eff}\ne 0$. Analogously, maximum in
the right half plane $\mu
>0$ (which corresponds to ($p=+1$)-solution \rf{s1}) is not the
extremum of $U_{eff}$. For completeness of picture, we also
included lines corresponding to the branch $q=-1$ (dashed lines in
Fig. \ref{pot}). The minimum of the right dashed line (for $X_{p}$
with $s=+1,\; p=-1,\; q=-1$) does not describe the extremum of
$U_{eff}$ because again $\partial_{\varphi} U_{eff}\ne 0$.

\section{\label{acceleration}cosmic acceleration and domain walls}

\setcounter{equation}{0}

Let us consider again the model with $d_1=D_0=4$ in order to
define stages of the accelerating expansion of our Universe. It
was proven that for certain conditions (see \rf{RR}-\rf{posit})
the effective potential $U_{eff}$ has local (for $z_0<z< 2$) or
global (for $z\geq 2$) positive minimum. The position of this
minimum is $\left(\varphi =0,\, \phi = (1/A)\ln
(1+sX^2_{p})\right)$ where $s=-1\, ,\; p=+1\, ,\; q=+1\,$ and
$z\in(z_{0},+\infty)$. Obviously, positive minimum of the
effective potential plays the role of the positive cosmological
constant. Therefore, the Universe undergoes the accelerating
expansion in this position. Thus, we can "kill two birds with one
stone": to achieve the stable compactification of the internal
space and to get the accelerating expansion of our external space.

We associate this acceleration with the late-time accelerating
expansion of our Universe. As it follows from eqs. \rf{RR} and
\rf{posit}, positive minimum takes place if the parameters are
positive and the same order of magnitude: $\Lambda_{eff} \sim R_1
\sim U(X) \sim h >0$. On the other hand, in KK models the size of
extra dimensions at present time should be $b_{(0)1} \lesssim
10^{-17}\mbox{cm} \sim 1\mbox{TeV}^{-1}$. In this case $R_1
\gtrsim b_{(0)1}^{-2} \sim 10^{34}\mbox{cm}^{-2}$. Thus, for the
TeV scale of $b_{(0)1} \sim 1$TeV we get that $\Lambda_{eff} \sim
R_1 \sim U(X) \sim h \sim 1\mbox{TeV}^2$. Moreover, in the case of
natural condition $\Lambda_D\sim \sqrt{|\mu|}$ we obtain that the
masses of excitations $m_{\varphi} \sim m_{\phi}\sim 1$TeV. The
above estimates clearly demonstrate the typical problem of the
stable compactification in multidimensional cosmological models
because for the effective cosmological constant we obtain a value
which is in many orders of magnitude greater than observable at
the present time dark energy $\sim 10^{-57}\mbox{cm}^{-2}$. The
necessary small value of the effective cosmological constant can
be achieved only if the parameters $R_1 \, , U(X) \, , h$ are
extremely fine tuned with each other to provide the observed small
value from equation $\Lambda_{eff}(X)=U_{0}(X)/3 -h$. We see two
possibilities to avoid this problem. Firstly, the inclusion of
different form-fields/fluxes may result in a big number of minima
(landscape) \cite{landscape1}-\cite{landscape4} with sufficient
large probability to find oneself in a dark energy minimum.
Secondly, we can avoid the restriction $R_1 \sim b_{(0)1}^{-2}
\sim 10^{34}\mbox{cm}^{-2}$ if the internal space is Ricci-flat:
$R_1 = 0$. For example, the internal factor-space $M_1$ can be an
orbifold with branes in fixed points (see corresponding discussion
in \cite{Zhuk2006}).

The WMAP three year data as well as CMB data are consistent with
wide range of possible inflationary models (see e.g.
\cite{WMAP2006}). Therefore, it is of interest to get the stage of
early inflation in our model. It is well known that it is rather
difficult to construct inflationary models from multidimensional
cosmological models and string theories. The main reason of it
consists in the form of the effective potential which is a
combination of exponential functions (see e.g. eq. \rf{1.7}).
Usually, degrees of these exponents are too large to result in
sufficiently small slow-roll parameters (see e.g. \cite{GZBR}).
Nevertheless, there is a possibility that in the vicinity of
maximum or saddle points the effective potential is flat enough to
produce the topological inflation \cite{topinfl,SSTM,R4}. Let us
investigate this possibility for our model.

As stated above, the value $\varphi=0$ corresponds to the internal
space value at the present time. Following this statement, we
found the minimum of the effective potential at this value of
$\varphi$. Obviously, the effective potential can also have
extrema at $\varphi \ne 0$. Let us investigate this possibility
for the model with $d_1=D_0=4$, i.e. for $c=0$. In this case, the
extremum condition of the effective potential reads
\ba{6.1}
\left.\partial_{\varphi}U_{eff}\right|_{\varphi_0,\phi_0}&=&
-\frac{1}{2}R_{1}(a+b)e^{(a+b)\varphi_0}+bU_{0}e^{b\varphi_0}\nn\\
&+& (ad_{1}+b)he^{(ad_{1}+b)\varphi_0}=0
 \ea
and
\begin{equation}\label{6.2}
\left.\partial_{\phi}U_{eff}\right|_{\varphi_0,\phi_0}=\left.
e^{b\varphi_0}\frac{\partial
U}{\partial\phi}\right|_{\phi_0}=0\;;\;\Longrightarrow\;\left.\frac{\partial
U}{\partial\phi}\right|_{\phi_0}=0\; .
\end{equation}
Here, $\varphi_0$ and $\phi_0$ define the extremum position. It
clearly follows from eq. \rf{6.2} that $\phi_0$ is defined by
equation $\partial U/\partial \phi =0$ which does not depend on
$\varphi$. Therefore, extrema of the effective potential may take
place only for $\phi_0$ which correspond to the solutions $X_{p}$
of eq. \rf{5.1} and different possible extrema should lie on the
sections $X_{p}=const $. So, we take $X_{+}$ (with $s=-1$ and
$q=+1$) which defines the minimum of $U_{eff}$ in the previous
section. Hence, $U_0$ in eq. \rf{6.1} is the same as for eq.
\rf{RR}.

\begin{figure}[htbp]
\centerline{\includegraphics[width=2.5in,height=2.5in]{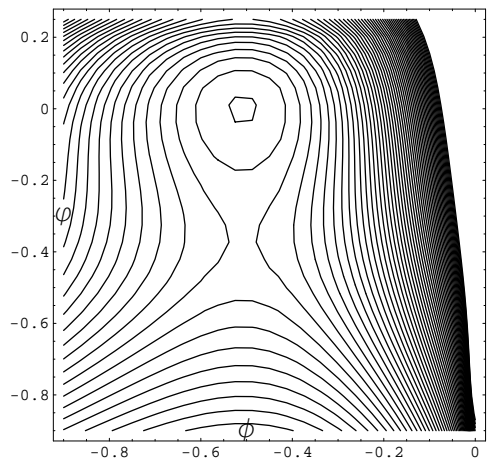}}
 \caption{Contour plot of the effective potential
$U_{eff}(\varphi ,\phi)$ for parameters
$z=1.99\;(\Lambda_{D}=1.99/2\;,\;|\mu|=1),\;D_{0}=d_{1}=4$ and
$h=U_{0}/6$. The rest of parameters follows from relations
\rf{RR}. We choose the branch corresponding to $s=-1,\, q=+1$.
This plot clearly shows the minimum and the saddle points of the
effective potential. \label{saddle}}
\end{figure}

Let us define now $\varphi_0$ from eq. \rf{6.1}. With the help of
inequalities \rf{J11>0} and \rf{posit} we can write $h=nU_{0}$
where $n\in (1/9,1/3)$. Taking also into account relations
\rf{RR}, eq. \rf{6.1} can be written as
\be{6.3} y^4 -\left(1+\frac{1}{3n}\right)y +\frac{1}{3n}=0\, , \ee
where we introduced the definition $y \equiv \exp{(a\varphi_0)} =
\exp{(\varphi_0/\sqrt{3})}$ and put $d_1=D_0=4$. Because $y=1$ is
the solution of eq. \rf{6.3}, the remaining three solutions
satisfy the following cubic equation:
\be{6.4} y^3+y^2+y-\frac{1}{3n}=0\, . \ee
It can be easily verified that the only real solution of this
equation is
\be{6.5} y_{0}=\frac{1}{3}\left(-1-2\nu+\frac{1}{\nu}\right)\; ,
\ee
where
\ba{6.6} \nu &=&\frac{2^{1/3}n}{\left(9n^2 +7n^3 +3\sqrt{9n^4 +14
n^6
+9n^6}\right)^{1/3}}\;;\;\nn\\n&\in&\left(\frac{1}{9},\frac{1}{3}\right)\;
. \ea
Thus $\varphi_{0}(y_0)$ and $\phi_0(X_{+})$ define new extremum of
$U_{eff}$. To clarify the type of this extremum we should check
signs of the second derivatives of the effective potential in this
point. First of all we should remember that in the case $c=0$
mixed second derivative disappears. Concerning second derivative
with respect to $\phi $, we obtain
\be{6.7} J_{22} \equiv \left.\frac{\partial^2 U_{eff}}{\partial
\phi^2}\right|_{\, \varphi_0,\phi_0} =
e^{b\varphi_0}\left.\frac{\partial^2 U}{\partial
\phi^2}\right|_{\, \phi_0} >0 \ee
because we got in previous section $\left.\partial^2 U/\partial
\phi^2\right|_{\, \phi_0(X_{+})} >0$. Second derivative with
respect to $\varphi $ reads
\ba{6.8} J_{11} &\equiv &\left.\frac{\partial^2 U_{eff}}{\partial
\varphi^2}\right|_{\, \varphi_0,\phi_0} = y_0^2
U_0\left[-6\left(\frac13 +n\right)y_0 + \frac43 \right.\nn\\&+&
\left.12ny_0^4\right] = 2y_0^2U_0\left[(3n+1)y_0 -
\frac43\right],\ea
where we took into account eq. \rf{6.3}. Simple analysis shows
that $(3n+1)y_0 < 4/3$ for $n\in (1/9,1/3)$. Keeping in mind that
$U_0>0$ we obtain $J_{11}<0$. Therefore, our extremum is the
saddle surface\footnote{Similar analysis performed for the branch
with $s=+1,\, q=-1$ (right dashed line in the Fig. \ref{pot})
shows the existence of the global negative minimum with $\varphi
\ne 0$ along the section $X_{-}=const$.}. Figure \ref{saddle}
demonstrates contour plot of the effective potential in the
vicinity of the local minimum and the saddle point.

Therefore, we arrived at very interesting  possibility for the
production of an inflating domain wall in the vicinity of the
saddle point. The mechanism for the production of the domain walls
is the following \cite{topinfl}. If the scalar field $\varphi$ is
randomly distributed, some part of the Universe will roll down to
$\varphi=0$, while in others parts it will run away to infinity.
Between any two such regions there will appear domain walls. In
Ref. \cite{SSTM}, it was shown for the case of a double-well
potential $V_{dw}(\varphi) = (\lambda/4)(\varphi^2-\xi^2)^2$ that
a domain wall will undergo inflation if the distance $\xi$ between
the minimum and the maximum of $V_{dw}$ exceeds a critical value
$\xi_{cr} =0.33M_{Pl}\, \to \, \kappa_0 \xi_{cr} = 1.65$. In our
case it means that the distance $|\varphi_0|$ between the local
minimum and the saddle point should be greater than $\xi_{cr}$:
$\, |\varphi_0| \geq 1.65$. Unfortunately, for our model
$\varphi_0(n) < 1.65$ if $n\in (1/9,1.3)$. For example, in the
most interesting case $n\to 1/3 $ (where $ \Lambda_{eff} \to 0$
(see eq. \rf{RR})) we obtain $|\varphi_0| \to 1.055$ which is less
than $\xi_{cr}$. Moreover, our domain wall is not thick enough in
comparison with the Hubble radius. The ration of the
characteristic thickness of the wall to the horizon scale is given
by $r_wH \approx \left|U_{eff}/3\partial_{\varphi\varphi}
U_{eff}\right|^{1/2}_{\varphi_0,\phi (X_+)} \to 0.454$ for $n\to
1/3$ which is less than the critical value 0.48 for a double-well
potential. Thus, here there is no a sufficiently large (for
inflation) quasi-homogeneous region of the energy density. Our
potential is too steep. Obviously, the slaw roll parameter
$\epsilon\approx(1/2)\left(\partial_{\varphi}U_{eff}/U_{eff}\right)^{2}_{\varphi_{0},\phi_{0}}$
is equal to zero in the saddle point. However, another slow roll
parameter
$\eta\approx\left|\partial^{2}_{\varphi\varphi}U_{eff}/U_{eff}\right|
_{\varphi_{0},\phi_{0}}\to -1.617$ for $n\to 1/3$. Therefore, our
domain walls do not inflate in contrast to the case $R^4$ in Ref.
\cite{R4}.

\begin{figure}[htbp]
\centerline{\includegraphics[width=3in,height=2in]{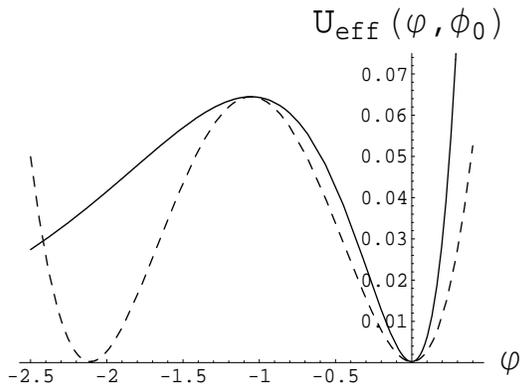}}
 \caption{Comparison of the potential $U_{eff}(\varphi ,\phi_0)$
 with a double well potential for parameters
$z=1.99\;(\Lambda_{D}=1.99/2\;,\;|\mu|=1),\;D_{0}=d_{1}=4$ and
$n=(1-0.001)/3$. \label{Maxhat}}
\end{figure}

In Fig. \ref{Maxhat} we present comparison between our potential
(solid line) and a double-well potential (dashed line) in the case
$n=(1-0.001)/3$. We see that our potential is flatter than a
double-well potential around the saddle point. However, our
calculations show that it is not enough for inflation.

\section{\label{conclu}Conclusions and discussions}

\setcounter{equation}{0}

We have shown that positive minimum of the effective potential
plays the double role in our model. Firstly, it provides the
freezing stabilization of the internal spaces which enables to
avoid the problem of the fundamental constant variation in
multidimensional models (\cite{Zhuk(IJMP)}-\cite{BZ}). Secondly,
it ensures the stage of the cosmic acceleration. However, to get
the present-day accelerating expansion, the parameters of the
model should be fine tuned. Maybe, this problem can be resolved
with the help of the idea of landscape of vacua
(\cite{landscape1}-\cite{landscape4}). We intend to investigate
this possibility in our forthcoming paper.

We have additionally found that our effective potential has the
saddle point. It results in domain walls which separates regions
with different vacua in the Universe. These domain walls do not
undergo inflation because the effective potential is not flat
enough around the saddle point.

It is worth of noting that minimum in Fig. 3 (left solid line) is
metastable. In other words, classically it is stable but there is
a possibility for quantum tunnelling both in $\phi$ and in
$\varphi$ directions (see Fig. 4). We can avoid this problem in
$\phi$ direction in the case of parameters $z\ge 2$ (see footnote
7). However, tunnelling in $\varphi$ direction (through the
saddle) is still valid because $U_{eff}(\varphi ,\phi_0) \approx
e^{b\varphi} U(\phi_0) \rightarrow 0$ for $\varphi \rightarrow
-\infty$ which is less than any positive $\Lambda_{eff}$. It may
result in the materialization of bubbles of the new phase in the
metastable one (see e.g. \cite{Rubakov}). Thus, late-time
acceleration is possible only if characteristic lifetime of the
metastable stage is greater than the age of the Universe. Careful
investigation of this problem (including gravitational effects) is
rather laborious task which needs a separate consideration. As we
mentioned in footnote 8, there is also the global negative minimum
for right dashed line in Fig. 3 (it corresponds to the point
$(\varphi =0.67,\phi =1.66)$ for parameters taken in Fig. 3). This
minimum is stable both in classical and quantum limits. However,
the acceleration is absent because of its negativity.

Another very interesting feature of the model under consideration
consists in multi-valued form of the effective potential. As it
can be easily seen from eqs. \rf{1.7} and \rf{U}, for each choice
of parameter $\mu$ potential $U(\phi )$ (and consequently
$U_{eff}$) has two branches ($q=\pm 1$) which joint smoothly with
each other at $\phi=0$ (see Fig. \ref{pot}). It gives very
interesting possibility to investigate transitions from one branch
to another one by analogy with catastrophe theory or similar to
the phase transitions in statistical theory. However, as we
mentioned above, in our particular model the point $\phi =0$
corresponds to the singularity $\bar R, R \to \pm \infty$. Thus,
the analog of the second order smooth phase transition through the
point $\phi =0$ is impossible in our model. Nevertheless, there is
still a possibility for the analog of the first order transition
via quantum jumps from one branch to another one. In what follows,
we plan to investigate such "phase transitions" for non-linear
multidimensional models $f(R)$.

To complete the paper, we investigate some limiting cases.
Firstly, we consider the limit $h \to 0$ (for arbitrary $D_0$ and
$d_1$) where the form-fields are absent. From eqs. \rf{U0} -
\rf{ppUeff} we obtain the following system of equations:
\be{c1} R_1 = \frac{2b}{a+b} U_0(X)\, , \quad \left.
U_{eff}\right|_{extr} = \frac{a}{a+b}U_0(X) \ee
and
\ba{c2} J_{11} &=& -ab\, U_0(X)\, , \quad J_{21} = 0\, ,\;
\nn\\J_{22} &=&
\left[B(A-B)-\frac{sAB}{2X^2}(1+sX^2)\right]U_0(X)\, .\qquad \ea
Since for minimum should hold true the condition $J_{11}>0$, we
arrive at the conclusion: $R_1,\, U_0, \, U_{eff} <0$.
Consequently, the minimum of the effective potential as well as
the effective cosmological constant is negative and accelerating
expansion is absent in this limit. Therefore, the presence of the
form-fields is the necessary condition for the acceleration of the
Universe in the position of the freezing stabilization of the
internal spaces. Additionally, it can be easily seen that the
extremum position equation takes the same form as \rf{5.1}. Simple
analysis show that minimum takes place for the branch: $s=+1$
(i.e. $\mu>0$), $p=-1, \, q=-1$ and $z\in (-\infty ,+\infty )$. If
additionally we demand $z \to 0$ (i.e. $\Lambda_D \to 0 $ and
$\mu$ is fixed) then we reproduce the results of Ref. \cite{GZBR}.


\bigskip
{\bf Acknowledgments}

We are grateful to Alex Vilenkin for his very useful comments
concerning the fine tuning problem and its possible resolution
with the help of the idea of landscape.



\begin{thebibliography}{99}
\bibitem{WMAP2006}
D. N. Spergel et al, {\it Wilkinson Microwave Anisotropy Probe
(WMAP) Three Year Results: Implications for Cosmology},
astro-ph/0603449.
\bibitem{R^{-1}}
S. Capozziello, S. Carloni and A. Troisi, {\it Quintessence
without scalar field},"Recent Research Developments in Astronomy
and Astrophysics"-RSP/AA/21-2003, astro-ph/0303041; S.M. Carroll,
V. Duvvuri, M. Trodden and M.S. Turner,  Phys.Rev. D{{\bf 70}},
(2004), 043528, astro-ph/0306438; D.N. Vollick, Phys. Rev. D{{\bf
68}}, (2003), 063510, astro-ph/0306630; R. Dick, Gen. Rel. Grav.
{{\bf 36}}, (2004), 217, gr-qc/0307052; A.D. Dolgov and M.
Kawasaki, Phys. Lett. B{{\bf 573}}, (2003), 1, astro-ph/0307285;
S. Nojiri and S.D. Odintsov, Phys. Rev. D{{\bf 68}}, (2003),
123512, hep-th/0307288; S. Nojiri and S.D. Odintsov,
Mod.Phys.Lett. A{{\bf 19}} (2004) 627, hep-th/0307288; T. Chiba,
Phys. Lett. B{{\bf 575}}, (2003), 1, astro-ph/0310045; X. Meng and
P. Wang, Class. Quant. Grav. {{\bf 20}}, (2003), 4949,
astro-ph/0307354; G. M. Kremer and D. S. M. Alves, Phys.Rev.
D{{\bf 70}}, (2004), 023503, gr-qc/0404082; N. Furey and A.
DeBenedictis, Class.Quant.Grav. {{\bf 22}}, (2005), 313,
gr-qc/0410088; F. P. Schuller and M. N.R. Wohlfarth, Phys.Lett.
B{{\bf 612}}, (2005), 93, gr-qc/0411076; S. Nojiri and S.D.
Odintsov, {\it Introduction to modified gravity and gravitational
alternative for dark energy}, hep-th/0601213; L.Amendola,
D.Polarski and S.Tsujikawa, {\it Are $f(R)$ dark energy models
cosmologically viable?}, astro-ph/0603703; E.J. Copeland, M.Sami,
S.Tsujikawa, Int.J.Mod.Phys. D{{\bf 15}}, (2006), 1753,
hep-th/0603057; Anthony W. Brookfield, Carsten van de Bruck, Lisa
M.H. Hall, Phys.Rev. D{{\bf74}}, (2006), 064028, hep-th/0608015;
T.P. Sotiriou, Phys.Lett. B{\bf{645}}, (2007), 389, gr-qc/0611107.
\bibitem{new}
S. Nojiri and S.D. Odintsov, Phys. Lett. B{{\bf 576}}, (2003), 5,
hep-th/0307071;
\bibitem{Carroll}
S. M. Carroll, A. De Felice, V. Duvvuri, D. A. Easson, Mark
Trodden and M. S. Turner, Phys.Rev. D{{\bf 71}}, (2005), 063513,
astro-ph/0410031.
\bibitem{Zhuk(IJMP)}
A. Zhuk, Int. Journ. Mod. Phys. D{{\bf 11}}, (2002), 1399,
hep-ph/0204195.
\bibitem{Kub} P. Loren-Aguilar, E. Garcia-Berro, J. Isern, and
Yu.A. Kubyshin, Class. Quant. Grav. {\bf{20}}, (2003), 3885,
astro-ph/0309723 .
\bibitem{GSZ}U. G\"unther, A. Starobinsky and A. Zhuk,  Phys. Rev.
D {\bf{69}}, (2004), 044003, hep-ph/0306191.
\bibitem{BZ}
V. Baukh and A. Zhuk, Phys. Rev. D{{\bf 73}}, (2006), 104016,
hep-th/0601205.
\bibitem{Alimi}
J.-M. Alimi, V.D. Ivashchuk, S.A. Kononogov and V.N. Melnikov,
Gravitation \& Cosmology, {\bf 12}, (2006), 173, gr-qc/0611015.
\bibitem{Uzan}
J.-P. Uzan, Rev. Mod. Phys. {{\bf 75}}, (2003), 403,
hep-ph/0205340.
\bibitem{GZ(PRD1997)}
U. G\"unther and A. Zhuk, Phys. Rev. D {\bf{56}}, (1997), 6391,
gr-qc/9706050.
\bibitem{GZBR}
U. G\"unther, A. Zhuk, V.B. Bezerra and C. Romero, Class. Quant.
Grav. {\bf{22}}, (2005), 3135, hep-th/0409112.
\bibitem{GMZ(PRDb)}
U. G\"unther, P. Moniz and A. Zhuk, Phys. Rev. D {\bf {68}},
(2003), 044010, hep-th/0303023.
\bibitem{GMZ(PRDa)}
U. G\"unther, P. Moniz and A. Zhuk, Phys. Rev. D {\bf{66}},
(2002), 044014, hep-th/0205148.
\bibitem{GMZ(ASS)}
U. G\"unther, P. Moniz and A. Zhuk, Astrophysics and Space Science
{\bf{283}}, (2003), 679, gr-qc/0209045.
\bibitem{FR} P.G.O. Freund and M.A. Rubin, Phys. Lett. B
{\bf{97}}, (1980), 233.
\bibitem{GZ(PRD2000)}
U. G\"unther and A. Zhuk, Phys. Rev. D {\bf{61}}, (2000), 124001,
hep-ph/0002009.
\bibitem{RZ} M. Rainer and A. Zhuk, Phys. Rev. D {\bf{54}},
(1996), 6186, gr-qc/9608020.
\bibitem{Ivashchuk}
V.D.Ivashchuk and V.N.Melnikov, Class.Quant.Grav. {\bf{12}},
(1995) 809, gr-qc/9407028; A.Zhuk, Class.Quant.Grav. {\bf{13}},
(1996) 2163; U.Gunther and A.Zhuk, Class.Quant.Grav. {\bf{15}},
(1998) 2025, gr-qc/9804018.
\bibitem{landscape1}
R. Bousso and J. Polchinski, JHEP {{\bf 0006}}, (2000), 006,
hep-th/0004134.
\bibitem{landscape2}
L. Susskind, The anthropic landscape of string theory;
hep-th/0302219. 
\bibitem{landscape3}
R. Kallosh and A. Linde, JHEP {{\bf 0412}}, (2004), 004,
hep-th/0411011.
\bibitem{landscape4}D. Schwartz-Perlov and A. Vilenkin, JCAP {{\bf
0606}}, (2006), 010, hep-th/0601162.
\bibitem{Zhuk2006}
A. Zhuk, Conventional cosmology from multidimensional models;
hep-th/0609126.
\bibitem{topinfl}
A. Linde, Phys. Lett. B {{\bf 327}}, (1994) 208,
hep-th/9402031; A. Vilenkin,  Phys. Rev. Lett. {{\bf 72}}, (1994),
3137, hep-th/9402085.
\bibitem{SSTM}
 N. Sakai, H. Shinkai, T. Tachizawa and K. Maeda, Phys.Rev. D {{\bf 53}}, (1996) 655; Erratum-ibid.
 D {{\bf 54}}, (1996) 2981, gr-qc/9506068.
\bibitem{R4}
 J. Ellis, N. Kaloper, K. A. Olive and J. Yokoyama, Phys.Rev. D {{\bf 59}}, (1999)
 103503, hep-ph/9807482.
\bibitem{Rubakov}
V.A. Rubakov and S.M. Sibiryakov, Theor.Math.Phys., {\bf{120}}
(1999), 1194, gr-qc/9905093.


\end{thebibliography}
\end{document}